\documentclass{article}

\usepackage{PRIMEarxiv}

\usepackage[utf8]{inputenc} 
\usepackage[T1]{fontenc}    
\usepackage{hyperref}       
\usepackage{url}            
\usepackage{booktabs}       
\usepackage{amsfonts}       
\usepackage{nicefrac}       
\usepackage{microtype}      
\usepackage{lipsum}
\usepackage{fancyhdr}       
\usepackage{graphicx}       
\usepackage{framed}
\usepackage{xcolor}
\usepackage{colortbl}
\graphicspath{{media/}}     

\title{Can a Chatbot Support Exploratory Software Testing? Preliminary Results
}

\author{
  Rubens Copche \\
  Grupo TCM \\
  Assis, Brazil \\
  \texttt{rcopche@gmail.com} \\
   \And
  Yohan Duarte Pessanha \\
  Federal University of Sao Carlos (UFSCar) \\
  São Carlos, Brazil \\
  \texttt{yohandpessanha@gmail.com} \\
  \And
  Vinicius Durelli \\
  Federal University of Sao Joao Del Rei (UFSJ) \\
  Sao Joao Del Rei, Brazil \\
  \texttt{durelli@ufsj.edu.br} \\
  \And
  Marcelo Medeiros Eler \\
  University of Sao Paulo (USP) \\
  Sao Paulo, Brazil \\
  \texttt{marceloeler@usp.br} \\
  \And
  Andre Takeshi Endo \\
  Federal University of Sao Carlos (UFSCar) \\
  Sao Carlos, Brazil \\
  \texttt{andreendo@ufscar.br} \\
}

\begin{document}
\maketitle

\begin{abstract}
Tests executed by human testers are still widespread in practice and fill the gap left by limitations of automated approaches. Among the human-centered approaches, exploratory testing is the \textit{de facto} approach in agile teams. Although it is focused on the expertise and creativity of the tester, the activity of exploratory testing may benefit from support provided by an automated agent that interacts with the human testers. 
     This paper presents a chatbot, called BotExpTest, designed to support testers while performing exploratory tests of software applications. 
     We implemented BotExpTest on top of the instant messaging social platform Discord; this version includes functionalities to report bugs and issues, time management of test sessions, guidelines for app testing, and presentation of exploratory testing strategies. 
     To assess BotExpTest, we conducted a user study with six software engineering professionals. They carried out two sessions performing exploratory tests along with BotExpTest. 
     Participants were capable of revealing bugs and found the experience to interact with the chatbot positive. 
     Preliminary analyses indicate that chatbot-enabled exploratory testing may be as effective as similar approaches and help testers to uncover different bugs. 
     Bots are shown to be valuable resources for Software Engineering, and initiatives like BotExpTest may help to improve the effectiveness of testing activities like exploratory testing.
\end{abstract}

\keywords{software testing \and bots \and exploratory tests}

\section{Introduction}

Due to their practical use and broad applicability, 
a myriad of bots that vary in complexity have been designed, developed, and deployed in widely varying contexts.
Over the last decade, 
technological advancements have enabled \emph{bots} to play an ever increasingly important role in many areas, 
particularly in software development. 
This emerging technology has garnered the interest of both software development researchers and practitioners, 
as bots can serve as human assistants for a variety of software development-related tasks.
In this particular context, 
bots that provide support for specific aspects of software development, 
such as keeping project-related dependencies up-to-date, 
are referred to as \emph{devbots}~\cite{erlenhov2019}. 

Recent developments in machine learning algorithms and natural language processing have led to 
the creation of bots that provide more user-friendly experiences. 
Bots that harness 
natural language processing capabilities to provide  more intuitive and user-friendly experiences are commonly referred to as \emph{chatbots}. 
As their name implies, 
chatbots are software programs designed to replicate human-like conversations or interactions with users~\cite{zadrozny2000natural,shawar2007chatbots}. 

As mentioned, 
bots have been utilized to support various software engineering tasks~\cite{storey2016disrupting,paikari2019chatbot,sharma2019smart,erlenhov2020,okanovic2020can}. 
We set out to examine how chatbots can be leveraged to assist testers throughout the testing process. 
Specifically, 
we posit that chatbots are well-suited for providing assistance to testers throughout the execution of Exploratory Testing (ET) tasks.
ET is an approach to software testing that entails carrying out a series of undocumented testing sessions to 
uncover faults.
ET leverages the skills and creativity of testers 
while they explore the system under test (SUT), 
and
the knowledge gained during ET sessions is then used to further refine the exploration. 
Hence, 
ET is a goal-focused, streamlined approach to testing that allows for flexibility in test design 
and keeps testers engaged throughout the testing process~\cite{bach2003exploratory,lyndsay2003adventures,kaner1993testing,souza2019exploratory}. 
Owing to these benefits, 
ET has been gaining traction as a complement to fully scripted testing strategies~\cite{istqb2018}: 
when combined with automated testing, 
ET has the potential to increase test coverage and uncover edge cases. 
In fact, 
there is evidence suggesting that ET can be equally or even more effective than scripted testing in practical situations~\cite{ghazi2017}.

In practice, 
before ET sessions, testers engage with other testers and developers to gather project-related information. 
However, 
due to the complexity of most software projects, 
it becomes impractical to collect all relevant information beforehand. 
As a result, 
interruptions that arise during ET sessions for the purpose of gathering additional information can disrupt the flow.
One potential solution to overcome this issue is to employ a chatbot that assists testers during ET sessions, 
providing guidance on the selection of input data for achieving different levels of exploration. 
Furthermore, 
the chatbot can encourage critical thinking and enable testers to make informed decisions.
To the best of our knowledge, 
this research is the first foray into the potential of a 
chatbot in maximizing the effectiveness of ET. 

This paper introduces \textsf{BotExpTest}, 
a chatbot designed to assist testers during ET sessions. 
\textsf{BotExpTest} was built on top of the Discord platform and includes features tailored to managing ET sessions and reporting bugs and issues. 
Additionally, 
it incorporates features aimed at enhancing testers' ability to gain further insights that can be utilized to delve further into the exploration of the SUT.\footnote{The development and evaluation of the current version of BotExpTest took place prior to the release of ChatGPT 
and other large language models (LLMs). However, in future work, we delve into the potential integration of these advanced technologies.} 
To evaluate how \textsf{BotExpTest} performs ``in the wild'', 
we conducted a user study with six practitioners. 
The results from the user study would seem to indicate that \textsf{BotExpTest} was able to help the participants to uncover several bugs. 
Moreover, 
the participants expressed a positive opinion about the experience and held an optimistic view regarding the potential future adoption of the tool.

\section{Related Work}
\label{sec:related}

\emph{Chatbots} are becoming increasingly popular in the software development domain because they can be very versatile. 
In this context, \emph{bots} are frequently classified based on their capacity of supporting different activities such as code review, tests, bug fixing, verification and deployment~\cite{storey2016disrupting}.  
Storey et al.~\cite{storey2020botse} surveyed developers and researchers to identify in which situations they use \emph{bots} to support software engineering activities. Here is what they found: 
to search and to share information, 
to extract and to analyze data, 
to detect and to monitor events,
to communicate in social media, 
to connect stakeholders and developers,
to provide feedback, and 
to recommend individual or collaborative tasks associated with software development.

Many studies have proposed \emph{bots} to support software development activities. 
\emph{Performobot} is a chatbot-based application that helps in planning, executing and reporting the results of tasks related to load and performance testing~\cite{okanovic2020can}. 
\emph{Smart Advisor} is an intelligence augmentation \emph{bot} that helps developers with project specifics by employing domain and knowledge modeling and in-process analytics to automatically provide important insights and answer queries using a conversational and interactive user interface~\cite{sharma2019smart}.
\emph{Repairnator} is a program repair \emph{bot} that creates software patches and provides an explanation for each bug fixed using natural language as a human collaborator would do~\cite{monperrus2019explainable}.
\emph{Tutorbot} uses machine learning to retrieve relevant content, guiding software engineers in their learning journey and helping them keep pace with technology changes~\cite{subramanian2019tutorbot}. 

To the best of our knowledge, 
there is no specific chatbot devised to help testers to conduct exploratory testing. 
In fact, 
a study conducted in Estonia and Finland found out that only 25\% of the software testing professionals apply ET with some tool support. 
Mind mapping tools are the most frequently used software, 
but testers also use text editors, spreadsheets and even actual pen and paper, in addition to checklists and paper notes (e.g. post-its)~\cite{pfahl2014exploratory}. 
In this context, 
Copche et al.~\cite{copche2021} introduced a specific kind of mind map called \textit{opportunity map} (OM) as a way to improve the ET of mobile apps. The authors conducted a study that compares OM-based ET with a traditional session-based approach (baseline). 

There are some tools to directly support activities involved in ET. 
For instance, Leveau et al.~\cite{leveau2020fostering} designed a tool called \emph{Test Tracker} to prevent testers from running tests that have already been executed so they can run more diversified test sessions to further explore the SUT.

There have been some attempts to integrate ET with automated approaches. 
For instance, Shah et al.~\cite{shah2014towards} proposed an hybrid approach that combines the strengths of ET and scripted testing (ST). Broadly, they identified the weaknesses of ST and proposed to use the strengths of ET as a solution. Similarly, they identified the weaknesses of ET and proposed to use the strengths of ST as a solution. When it comes to ST, for example, test case design quality depends on the test designer skills. 
Considering the strengths of ET, 
which includes the application of domain knowledge and the observation of the system behavior for rapid feedback, the hybrid process allows the testers to explore the SUT freely and to utilize their intuitions and experience in identifying defects before writing test scripts.

\section{BotExpTest}
\label{sec:botexptest}

\begin{figure}[h]
\centerline{\fbox{\includegraphics[width=0.8\columnwidth]{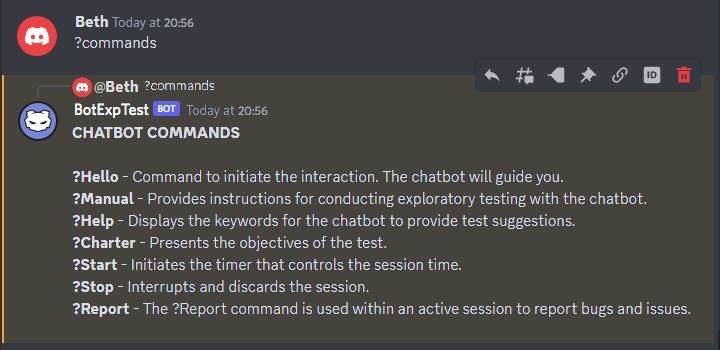}}}
\caption{BotExpTest replies its main commands.}
\label{fig:botcomand}
\end{figure}

This section presents the design and main features of a chatbot we developed to support ET. We set out by investigating existing work about ET, 
its core practices and envisioned how a chatbot would help the tester to conduct more effective ET sessions.
To validate these ideas, 
we implemented
\textsf{BotExpTest}.
Figure~\ref{fig:botcomand} shows the example of an interaction with BotExpTest: 
tester Beth types \textbf{?commands} and then \textsf{BotExpTest} shows all commands accepted. 

\subsection{Implementation}

As instant messaging platforms are widely adopted and are today an essential part of software projects, 
we opted to develop \textsf{BotExpTest} on top of them. 
For this first release, we settled on using the Discord platform\footnote{\url{https://discord.com}}.
Discord provides an open source platform with highly configurable features for users and bots. 
\textsf{BotExpTest} is implemented as a Node.js project; it has 22 classes and around 1.3K lines of JavaScript code. 
It takes advantage of the Discord API to capture interactions from testers in the chat, 
as well as to generate its own messages. 
To make the ET process auditable, all messages exchanged between the chatbot and the testers are recorded in a MongoDB database.
 \textsf{BotExpTest} is available as an open source project at:
 
\begin{center}
\url{https://github.com/rcopche/BotExpTest}   
\end{center}

\begin{figure}[h]
\centerline{\fbox{\includegraphics[width=0.6\columnwidth]{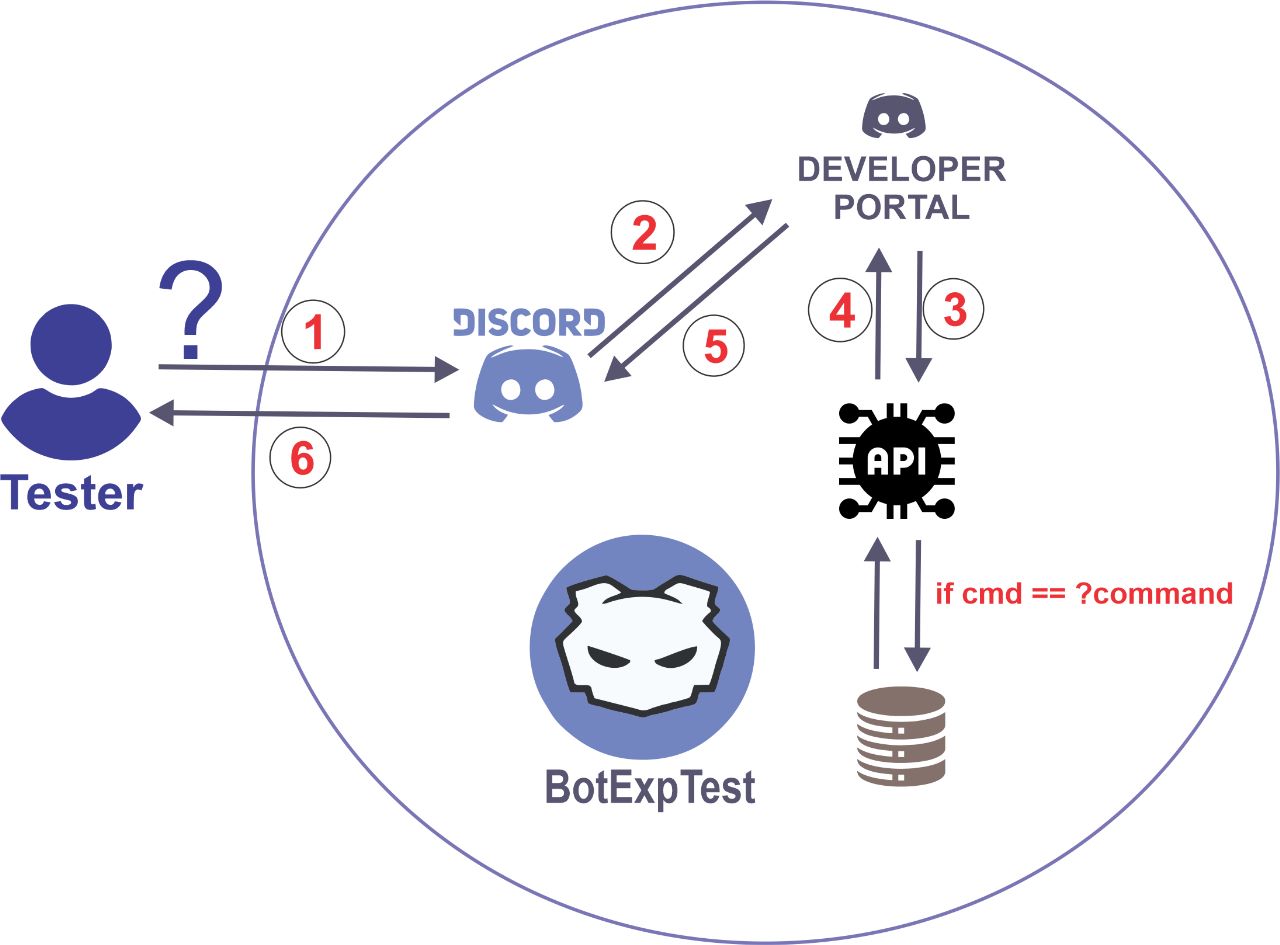}}}
\caption{Architecture of \textsf{BotExpTest}.}
\label{fig:arquiteturabot}
\end{figure}

Figure~\ref{fig:arquiteturabot} presents an overview of the \textsf{BotExpTest} architecture.
The interaction starts with the tester writing a message (command) to \textsf{BotExpTest} via Discord (step~1). 
The message passes through the Discord Developer Portal, which is then accessible by means of an API (steps~2-3). 
Up to this point, \textsf{BotExpTest} interprets the message typed by the tester and reacts by sending a reply (steps~4-5). 
Finally, the \textsf{BotExpTest}'s response is shown to the tester and new interactions may occur. 
\textsf{BotExpTest} may also be the one that starts the interaction.

\subsection{Main Features} 
\label{chp:funcionalidades}

During the first interaction between tester and \textsf{BotExpTest}, 
the first reaction of the chatbot is to present itself, 
giving some pieces of information about how to perform the next steps in the ET session. 
As a convention, 
testers begin an interaction with \textsf{BotExpTest} using a message that starts with '?'. 
Adhering to this convention can be beneficial in scenarios where several testers are communicating with each other in a chat, 
as it indicates when testers intend to engage with the bot. 
For this version, 
the features implemented by \textsf{BotExpTest} were elicited, prioritized and implemented; they are described next.

\begin{figure}[h]
\centerline{\fbox{\includegraphics[width=0.9\columnwidth]{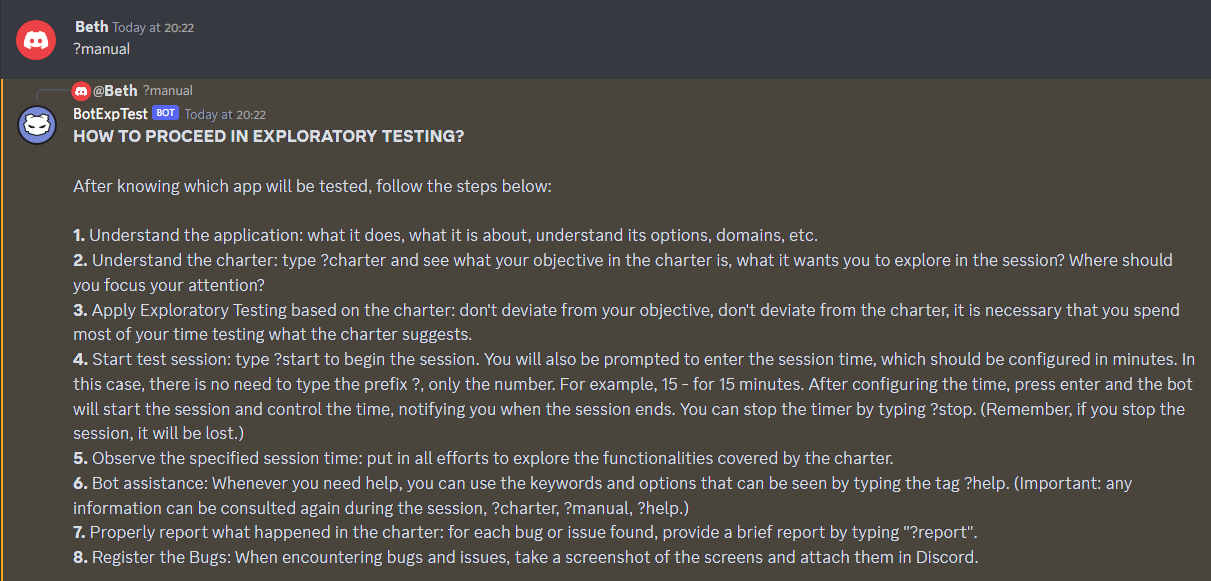}}}
\caption{\textsf{BotExpTest} replies to ?manual.}
\label{fig:botmanual}
\end{figure}

\textbf{Description of the test procedure:}
Using command \textbf{?manual}, \textsf{BotExpTest} shows a step-by-step description about how the test sessions are organized and should be conducted, as well as the main features provided to tester by the chatbot. This is illustrated in Figure~\ref{fig:botmanual}. 

\textbf{Charters:} 
In ET, \textit{charters} are used to organize the tests and represent the goals that are supposed to be achieved in a test session. 
\textsf{BotExpTest} provides an interface to set up the test charters and are available to testers by using the message \textbf{?charter}. 
Besides the charter name, app name, and the goals description, it is also possible to attach images and other files related to the charter. 

\begin{figure}[h]
\centerline{\fbox{\includegraphics[width=0.6\columnwidth]{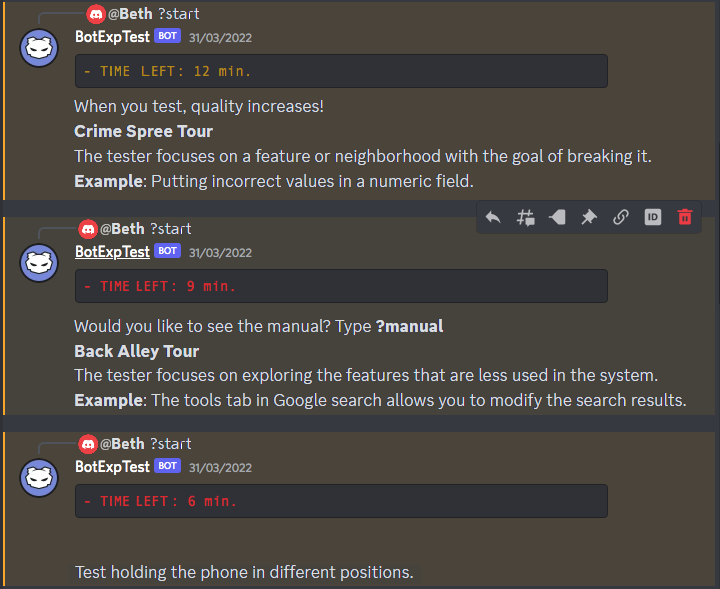}}}
\caption{Time alerts and suggestions.}
\label{fig:botIntAut}
\end{figure}

\textbf{Time management of testing sessions:}
In ET, testing sessions are conducted within a limited time frame; usually, testers need to keep track of time. 
As the tester is constantly interacting in the chat, \textsf{BotExpTest} reminds her about the remaining time in the session, from time to time.  
To signal the start of a session, the tester should type the command \textbf{?start}. 
\textsf{BotExpTest} then asks the time limit and starts to monitor the time elapsed in the session. 
During the time range of the session, \textsf{BotExpTest} keeps track of all interactions that occurred. Figure~\ref{fig:botIntAut} shows how the time alerts are presented to the tester. 

\begin{figure}[h]
\centerline{\fbox{\includegraphics[width=0.7\columnwidth]{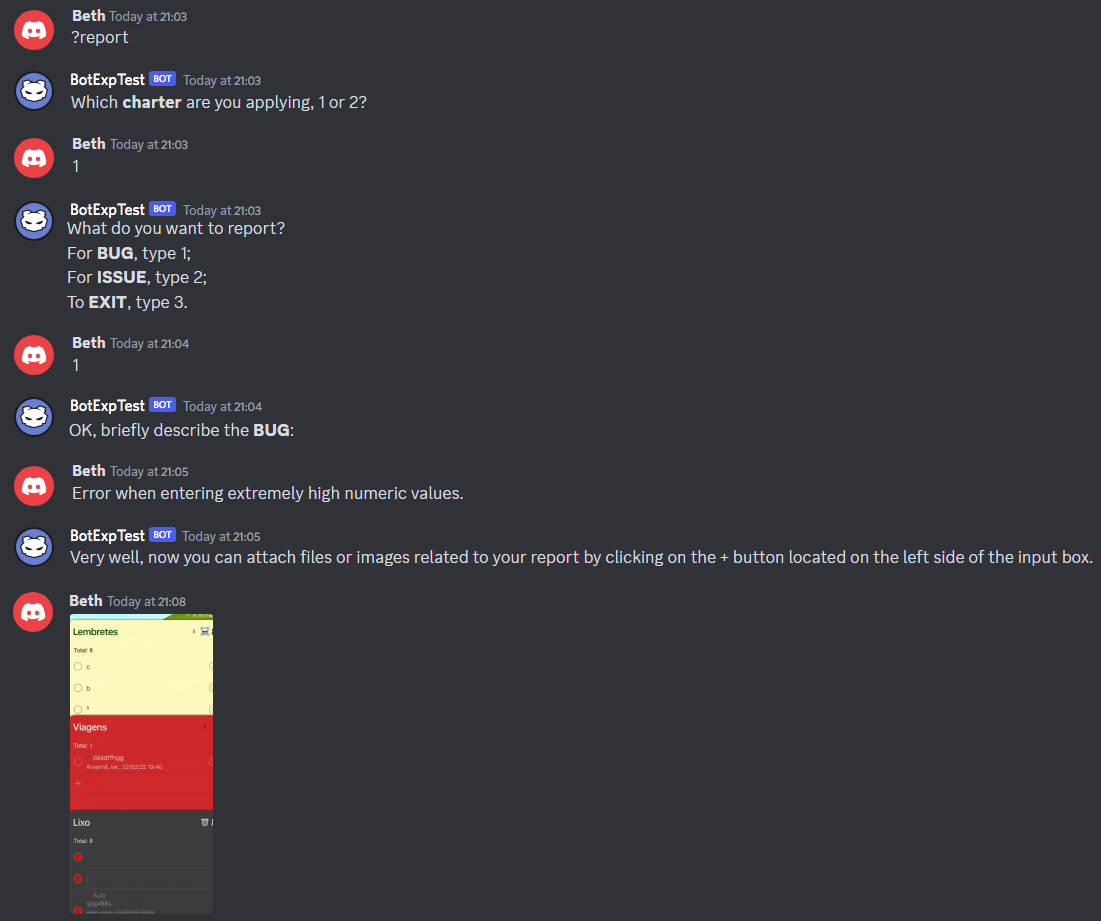}}}
\caption{Reporting a bug. }
\label{fig:botrelatar}
\end{figure}

\textbf{Bug and issue reporting:}
The identification of bugs and issues are the main outcomes of a test session. To avoid using other tools for this task, \textsf{BotExpTest} registers occurrences of bugs or issues. 
To do so, the tester types command \textbf{?report}; this task is exemplified in Figure~\ref{fig:botrelatar}.
The tester interacts with \textsf{BotExpTest} so that the charter, type (bug or issue), a detailed description and potential attachments (e.g., screenshots of the bug) are provided. 
The current version only stores the bug report, but it is possible in future to integrate it with external tools like GitHub, Jira or Azure DevOps.

\begin{figure}[t]
\centerline{\fbox{\includegraphics[width=0.8\columnwidth]{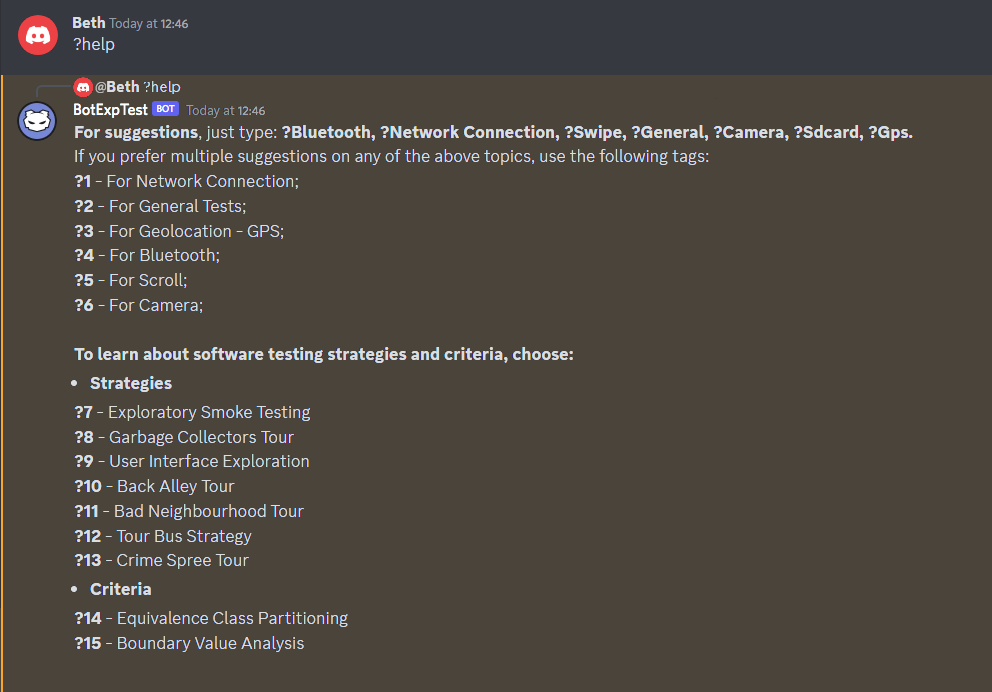}}}
\caption{Command ?help.}
\label{fig:botajuda1}
\end{figure}

\textbf{Curated knowledge about exploratory testing}: 
The idea of this feature is for \textsf{BotExpTest} to have a curated list of resources about the use of ET techniques. 
In future, \textsf{BotExpTest} could be fine-tuned for specific projects so that testers are better equipped to conduct exploratory tests. 
Figure~\ref{fig:botajuda1} shows the available resources after the tester types command \textbf{?help}. 
By typing one of the presented options, \textsf{BotExpTest} shows a detailed explanation about the concept and how to apply it during the tests. 
For some options, the chatbot replies with questions. 
We anticipate that this feature may help the testers to gain more insights and execute more effective tests. 

Currently, \textsf{BotExpTest} provides resources related to three main groups. 
Group~\textit{(i)} brings well-known testing criteria that may help the tester to design black-box tests. For example, classical criteria like equivalence partition and boundary-value analysis are presented. 
Group~\textit{(ii)} is composed of strategies for ET that are well-known and have been adopted in the literature~\cite{conf/icst/MicallefPB16,whittaker2009exploratory}. For example, \textit{Bad Neighborhood Tour} instructs the tester to revisit buggy parts of software since bugs tend to cluster together.
Finally, Group~\textit{(iii)} contains guidelines for mobile app testing. For example, there are several test scenarios related to specific characteristics of mobile apps like network connections, geolocation, Bluetooth, camera, and UI events (scrolling, swipe, etc).

\textbf{Active suggestions}:
During a test session, \textsf{BotExpTest} can actively start an interaction with the tester. 
The chatbot can present some piece of information obtained from the curated knowledge about exploratory testing. 
We believe that actively interacting with the tester would increase the engagement with the tests. 
For the current version, the decision about the timing and the information provided is made randomly. In future, we expect that \textsf{BotExpTest} could be evolved to make a more informed decision about the interaction needed at a specific point of time. 

\section{User Study}\label{sec:user-study}

This section describes an empirical study conducted with the aim to provide an initial evaluation of \textsf{BotExpTest}. 
To this end, 
we posed the following research questions (RQs):

\begin{itemize}
    \item RQ1: How is the interaction with \textsf{BotExpTest} during the exploratory testing?
    
    \item RQ2: How do the participants perform with respect to the detection of bugs?
    
    \item RQ3: How do the participants perceive \textsf{BotExpTest}?
\end{itemize}

\begin{figure}[h]
 
 \centering
  \includegraphics[width=0.8\columnwidth]{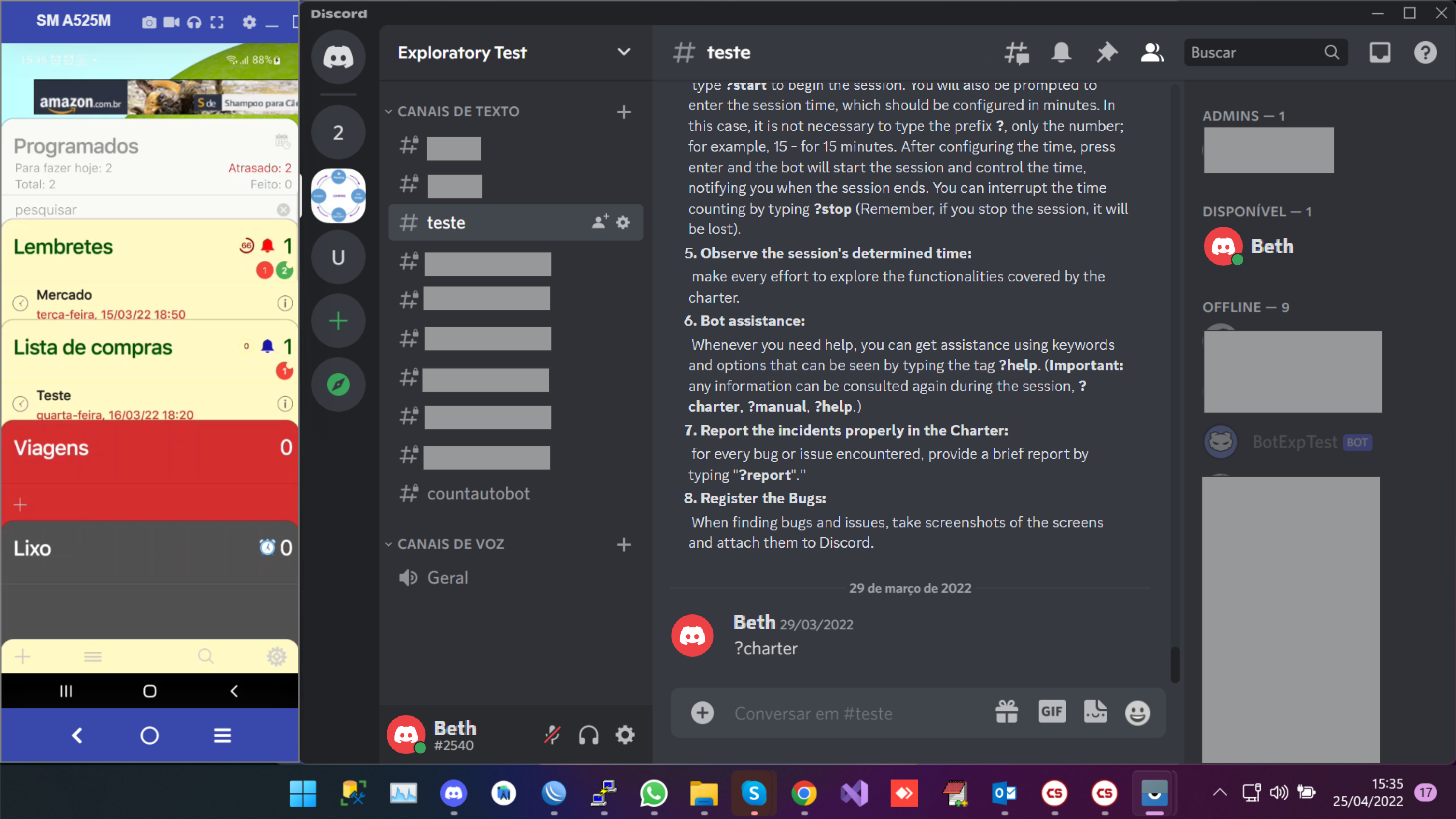}
  \caption{Setup used by the participants.}
  \label{fig:MainScreen}
\end{figure}

To answer these questions, 
we conducted a user study with six participants that were asked to use \textsf{BotExpTest} to support the ET of a mobile app named \textit{Reminders}\footnote{\url{https://play.google.com/store/apps/details?id=com.chegal.alarm}}.
All participants work in the industry and have experience with software development and testing. 
We adopted the app and related charters that were openly available from Copche~et~al.~\cite{copche2021}; the rationale here is to make some analyses concerning similar approaches\footnote{We used the same app version, charters and length of test sessions.}. 
To collect the needed data and observe the participants, we set up a computer with screen recording, and mirroring the mobile device running the app under test. 
Each participant was invited to use this computer in order to perform the tasks of the study.
Figure~\ref{fig:MainScreen} illustrates the testing environment used by the participants; the reminders app is shown on the left, while the Discord UI (along with \textsf{BotExpTest}) is presented on the right.

Initially, 
we provided the participants with detailed instructions about how to perform the testing tasks. 
They were instructed to strictly follow the charter, 
report any bug or issue identified, and the time management of the session was supported by \textsf{BotExpTest}.
All test sessions were recorded and the participants could think aloud about the tasks being carried out.
After the introduction, 
the study was divided into two parts:

\begin{itemize}
\item  \textbf{Training:} 
this session lasted approximately 15 minutes and served as an introduction to the usage of the chatbot. 
This allowed participants to become acquainted with BotExpTest and Discord, 
allowing them to experiment with possible interactions and explore the supported commands.

\item  \textbf{Test sessions:} this iteration took approximately half an hour, 
in which two test sessions with 15 minutes each occurred. Over the course of these two test sessions, 
the chatbot-enabled ET took place.
\end{itemize}

All data was then retrieved and analyzed. 
To answer RQ1, 
we looked at the interactions that occurred (messages exchanged) between the participant and \textsf{BotExpTest}. 
We called \textit{Active Interactions} the ones started by \textsf{BotExpTest}, 
while \textit{Reactive Interactions} are responses to the tester's inquiries.

As for RQ2, 
we analyzed and cataloged the bugs reported by participants. 
In particular, 
we cross-checked the bugs herein reported with the ones uncovered in Copche~et~al.~\cite{copche2021}. 
As an initial analysis, 
we intended to assess whether a chatbot-enabled ET can detect a different set of bugs with respect to similar approaches like baseline and OM (see Section~\ref{sec:related}).

We answered RQ3 with a Likert-scale survey that intends to understand the perception of participants. 
The questions are divided into \textit{(i)} how easy is to interact and use \textsf{BotExpTest}, \textit{(ii)} whether the user interface is adequate for the proposed functionalities, and \textit{(iii)} understanding the participants' perception about the effectiveness of chatbot-enabled ET.  
There is also an open question for comments and suggestions.  

\subsection{Analysis of Results}

\paragraph*{RQ1 - Interactions with BotExpTest.}
Table~\ref{tab:Interacoesgerais} shows the number of active and reactive interactions of both the chatbot and the participants; the values are also divided between training and test sessions. Overall, \textsf{BotExpTest} produced 581 interactions (121 in training and 460 in the test sessions), while the six participants had 496 interactions (144 in training and 352 in the test sessions). 

\begin{table}[h]
 \caption{Number of interactions (int.).}
 \centering
\begin{tabular}{|l|c|l|l|c|}
\hline
\multicolumn{2}{|c|}{\cellcolor{blue!15}\textbf{TRAINING}} & \cellcolor{blue!15} & \multicolumn{2}{c|}{\cellcolor{blue!15}\textbf{TEST SESSIONS}}\\
\hline
\multicolumn{2}{|c|}{\cellcolor{yellow!15}\textbf{BotExpTest}} & \cellcolor{blue!15} & \multicolumn{1}{c|}{\cellcolor{yellow!15}\textbf{BotExpTest}}&\\
\hline
Reactive int. & 107 &\cellcolor{blue!15}& Reactive int. & 340 \\ 
\hline
Active int. & 14 & \cellcolor{blue!15}& Active int. & 120 \\ 
\hline
\textcolor{red}{Total} & \textcolor{red}{121} & \cellcolor{blue!15}& \textcolor{red}{Total} & \textcolor{red}{460} \\ 
\hline
\multicolumn{2}{|c|}{\cellcolor{orange!15}\textbf{Participants}} & \cellcolor{blue!15} & \multicolumn{2}{c|}{\cellcolor{orange!15}\textbf{Participants}} \\
\hline
Reactive int. & 37 &\cellcolor{blue!15} & Reactive int. & 262 \\ 
\hline
Active int. (accepted) & 100 & \cellcolor{blue!15}& Active int. (accepted) & 90 \\ 
\hline
Active int. (invalid) & 7 & \cellcolor{blue!15}& Active int. (invalid) & 0 \\ 
\hline
\textcolor{red}{Total} & \textcolor{red}{144} & \cellcolor{blue!15}& \textcolor{red}{Total} & \textcolor{red}{352} \\ 
\hline
\end{tabular}
\label{tab:Interacoesgerais}
\end{table}

During training, \textsf{BotExpTest} was proportionally more reactive (reactive/active -- 107/14) and the participants were more active (37/100); they also typed 7 invalid commands. 
The most typed commands were related to the chatbot usage (?charter, ?commands, ?help, ?manual) and software testing resources. 
We observed that the participants were more active in the interactions because they were focused on figuring out how to use \textsf{BotExpTest}.

As for test sessions, \textsf{BotExpTest} was still more reactive (340/120) but proportionally had more active interactions. 
This occurred due to the reporting of bugs and issues (it asks actively for more pieces of information). 
This fact also impacted the participants' interactions: they were more reactive (262/90) and did not type any invalid command. 
The most typed commands were related to bug and issue reporting (e.g., ?report) and management of the test sessions (e.g., ?start, ?charter). 
We also observed that participants were focused on testing the app and this also limited the interactions with the chatbot. 

\begin{framed}\noindent
\textit{Answer to RQ1:} We observed reasonable interactions between the participants and \textsf{BotExpTest}. As the participants were exploring the chatbot in training, they had more active interactions. On the other hand, the interactions were more reactive and limited in the test sessions due to the time spent with exploratory testing of the app itself and reporting bugs and issues.
\end{framed}

\begin{figure}[h]
 \centering
  \includegraphics[width=0.5\columnwidth]{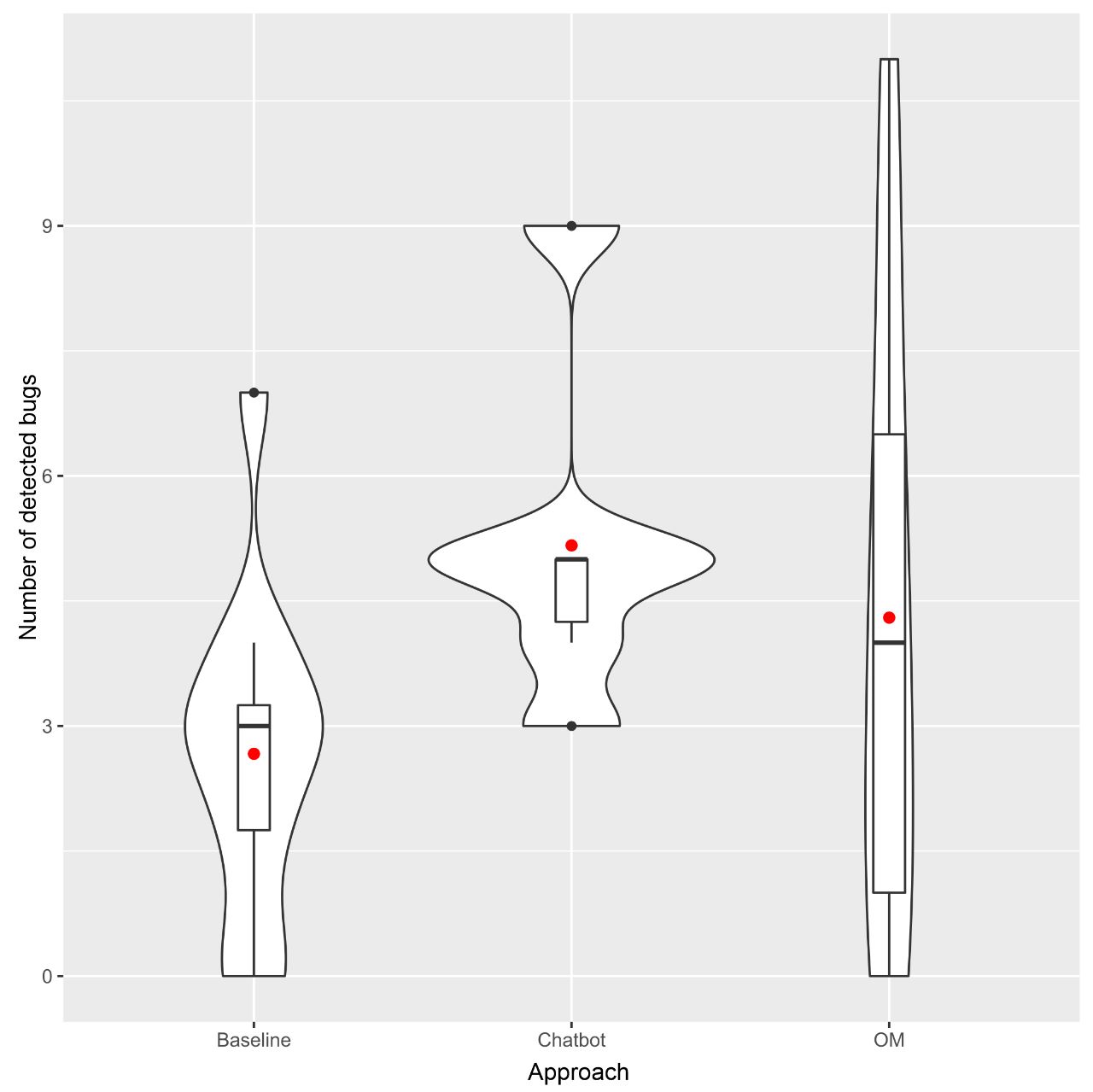}
  \caption{Number of bugs detected by the participants.}
  \label{fig:BlMoChat}
  \vspace{-.2cm}
\end{figure}

\paragraph*{RQ2 - Bugs.}
The six participants reported 31 bugs.
The most effective participant uncovered nine bugs, while one of them reported three bugs. On average, the participants detected 5.2 bugs (median: 5). The distribution of bugs detected per participant can be seen in Figure~\ref{fig:BlMoChat}, the boxplot/violin in the middle. 

Figure~\ref{fig:BlMoChat} also shows the results for the baseline and OM approaches~\cite{copche2021}. 
Observe that the average and median values of \textsf{BotExpTest} are slightly greater than baseline (avg: 2.7, median: 3) and OM (avg: 4.3, median: 4). 
The range of values is also smaller, so \textsf{BotExpTest} produced a more uniform performance of participants. 
Due to the differences of samples and participants' experience, these results may not be generalized. 
One may argue that the bug detection capability of \textsf{BotExpTest} is at least comparable to approaches without any support (i.e., baseline) or that adopted some supporting artifact (i.e., OM). 

\begin{figure}
 \centering
\includegraphics[width=0.35\columnwidth]{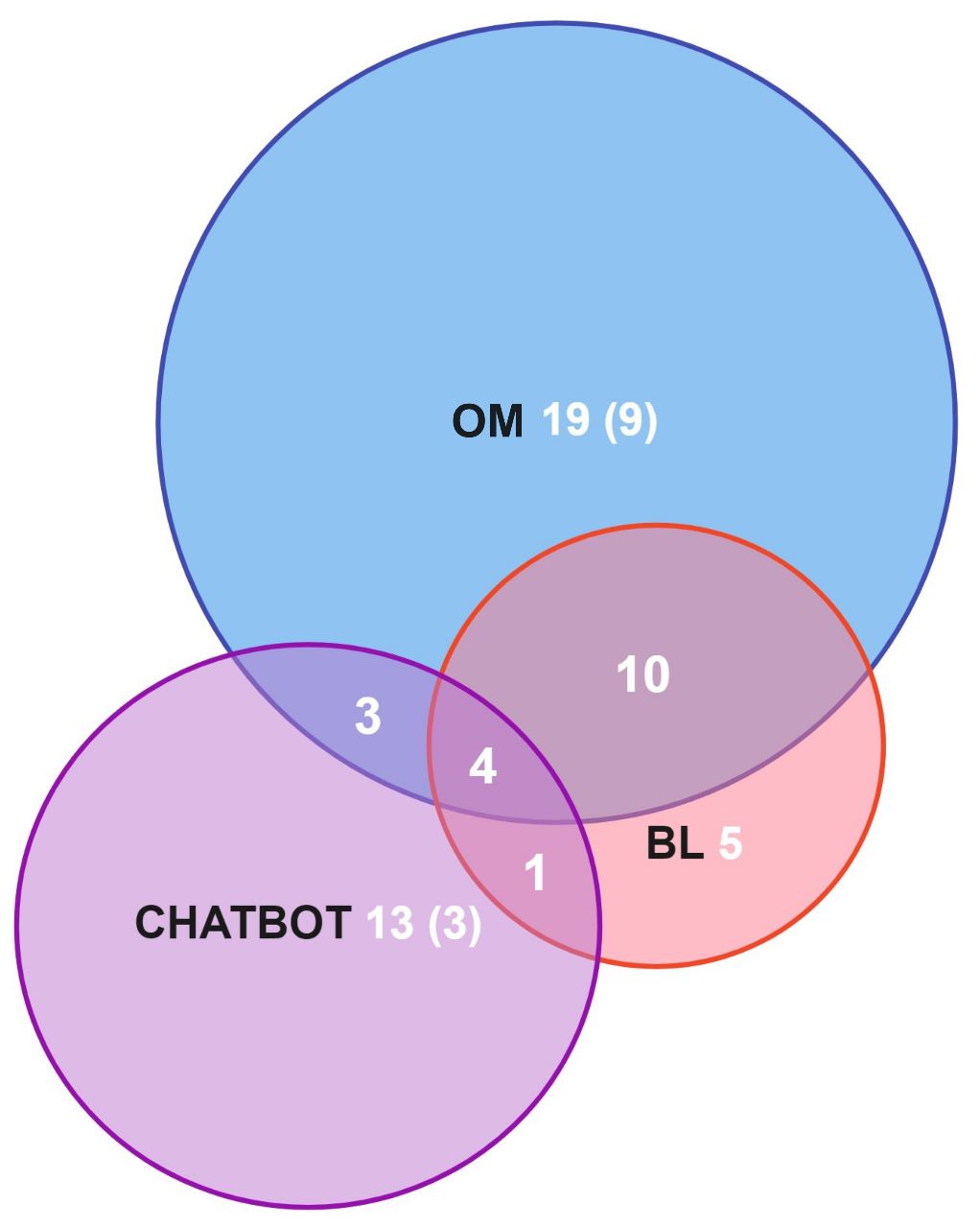}
\caption{Unique bugs. }
  \label{fig:venn}
  \vspace{-.5cm}
\end{figure}

Figure~\ref{fig:venn} shows a Venn diagram with the unique bugs detected by the approaches, 
numbers between parentheses are bugs tracked to specific suggestions of the approach.  
Out of 31 bugs reported by the participants using \textsf{BotExpTest}, 
21 were unique since some reported the same bug. 
Eight bugs have been uncovered in the Copche et al. study (1 by baseline, 3 by OM and 4 by both), 
but 13 yet-unknown bugs were detected in this study. 
We were able to map three out of these 13 bugs to specific insights provided by the chatbot. 

\begin{framed}\noindent
\textit{Answer to RQ2:} 
The participants were capable of uncovering an average of 5.2 bugs using \textsf{BotExpTest}. 
Their performance is comparable to similar approaches evaluated in the literature. 
Furthermore, 
\textsf{BotExpTest} supported the participants in detecting 13 previously unknown bugs. 
\end{framed}

\paragraph*{RQ3 - Participants' perception.}
For the part \textit{(i)} of questions, we asked about the easiness of finding information, whether proper instructions are provided, and usability in general. All participants agreed or strongly agreed that \textsf{BotExpTest} is easy to use and interact. 
As for part \textit{(ii)}, the questions asked about specific features like bug/issue reporting and time management of test sessions. Most participants strongly agreed or agreed that the user interface for those features is adequate. In particular, the bug/issue reporting was unanimously well-evaluated (all strongly agreed).

For the last part of questions, the responses indicated that participants would use a chatbot in similar tasks, and they perceived more organized test sessions. They also thought that \textsf{BotExpTest} helped them to find more bugs. 
We also asked if the chatbot helped them to understand new concepts of software testing, and whether the suggestions made by \textsf{BotExpTest} were helpful; all participants strongly agreed or agreed with those statements. 

From the open question and our observations, we draw the following thoughts. 
Participants believed that the chatbot helped to shorten the time spent with process tasks, saving more time to test the app. 
\textsf{BotExpTest} worked as a rich and centralized source of testing information; participants sometimes used the message history to revisit decisions and bugs detected. 
One suggested that it could support novice programmers testing their software, and another mentioned that it could help teams without QAs.
Finally, there were suggestions to add support for other testing tasks, like managing test scripts, tracking the status of test executions, and communication with stakeholders.

\begin{framed}\noindent
\textit{Answer to RQ3:} The participants perceived \textsf{BotExpTest} as a valuable resource while performing exploratory testing. Overall, the participants' perceptions were positive concerning the features, the ease of interaction, and testing resources.
\end{framed}

\section{Concluding Remarks}
\label{sec:concluding}

This paper presents an initial effort on using chatbots to support exploratory software testing. We implemented the first version of \textsf{BotExpTest} and evaluated it with six software development professionals. The results gave evidence that chatbot-enabled ET has potential to be as effective as similar approaches and received positive feedback from the participants. 

We recognize the limitations of this study, yet we intend to draw two main future initiatives from the preliminary findings obtained.  
First, future replications and extended controlled experiments are needed to better assess the impact of chatbots in software testing. 
Then, \textsf{BotExpTest} could be evolved with modern technologies so that it improves its testing support and interaction capabilities. 
One direction is to include the ability of observing the SUT (using e.g. monitoring or dynamic analyses as in~\cite{leveau2020fostering}). This ability would be used to feed the chatbot and provide more educated insights. 
On the human-bot interaction side, LLMs (like ChatGPT, Bard) and other similar technologies could be adopted to make the conversations more human-like, and provide a broader (and personalized) access to software testing knowledge.

\section*{Acknowledgments}

Andre T. Endo is partially supported by grant \#2023/00577-8, São Paulo Research Foundation (FAPESP). 
Yohan Pessanha is supported by grant \#2022/13469-6, São Paulo Research Foundation (FAPESP) and Coordenação de Aperfeiçoamento de Pessoal de Nível Superior (CAPES) grant 88887.801592/2023-00.

\bibliographystyle{unsrt}  
\bibliography{references}

\end{document}